\documentclass{emulateapj}

\usepackage{color}

\usepackage{epsfig}
\usepackage{rotate}

\def\ptsec{$^{\prime\prime}\mskip-7.6mu.\,$}

\shorttitle{10~$\mu$m Silicate Variability in TTS}
\shortauthors{Bary et al.}

\begin{document}

\title{Variations of the 10~$\mu$m Silicate Features in the Actively Accreting T~Tauri Stars: DG~Tau and XZ~Tau}

\author{Jeffrey S. Bary\altaffilmark{1}, Jarron M. Leisenring\altaffilmark{2}, Michael F. Skrutskie\altaffilmark{2}}
\altaffiltext{1}{Department of Physics and Astronomy, Colgate University, 13 Oak Drive,
Hamilton, NY  13346}
\altaffiltext{2}{Department of Astronomy, University of Virginia, P.O. Box 400325 
Charlottesville, VA 22904-4325}
\email{jbary@colgate.edu, jml2u@virginia.edu, mfs4n@virginia.edu}

\doublespace

\begin{abstract}

Using the Infrared Spectrograph aboard the \emph{Spitzer Space Telescope}, we observed multiple epochs of 11 actively accreting T~Tauri stars in the nearby Taurus--Auriga star forming region. In total, 88 low-resolution mid-infrared spectra were collected over 1.5 years in Cycles~2 and 3.  The results of this multi-epoch survey show that the 10~$\mu$m silicate complex in the spectra of two sources --- DG~Tau and XZ~Tau --- undergoes significant variations with the silicate feature growing both weaker and stronger over month- and year-long timescales.  Shorter timescale variations on day- to week-long timescales were not detected within the measured flux errors.  The time resolution coverage of this data set is inadequate for determining if the variations are periodic.  Pure emission compositional models of the silicate complex in each epoch of the DG~Tau and XZ~Tau spectra provide poor fits to the observed silicate features.  These results agree with those of previous groups that attempted to fit only single-epoch observations of these sources.  Simple two-temperature, two-slab models with similar compositions successfully reproduce the observed variations in the silicate features.  These models hint at a self-absorption origin of the diminution of the silicate complex instead of a compositional change in the population of emitting dust grains.  We discuss several scenarios for producing such variability including disk shadowing, vertical mixing, variations in disk heating, and disk wind events associated with accretion outbursts.
\end{abstract}


\section{Introduction}

The spatial distribution, size, and composition of silicate dust grains observed in circumstellar disks surrounding young pre-main-sequence stars result from a variety of physical and chemical evolutionary processes.  As they play a key role in the evolution of circumstellar disks toward planetary systems, an array of observational techniques has been employed to study the characteristics of these dust grains (sub-millimeter, millimeter, mid-infrared, polarization, etc., \cite[e.g.][]{wein1989,stro1989,beck1990,skru1990}.  Building upon much ground-based infrared photometry and imaging sensitive to the blackbody emission and scattered light from dust grains in circumstellar environments, the \emph{Spitzer} Infrared Spectrograph (IRS) has amassed an extensive catalog of near- to mid-infrared spectral energy distributions (SEDs) of young sources in the nearest star forming regions \citep{kess2005,furl2006,sarg2009,wats2009}.  Modeling of the SEDs of this large sample of T Tauri stars (TTS) ---  known to possess thick and actively accreting disks --- succeeds in defining the spatial and size distributions of the dust grains in these systems, potentially giving astronomers direct insight into the evolutionary status of these planetary building blocks.  For instance, SED models suggest the presence of `puffed-up' inner disk rims near the dust sublimation radius for some sources \citep{natt2001,dull2001,muze2003,muze2004} and constrain the sizes of the optically thin inner disk holes of transitional objects finding some to be as large as 40~AU \citep{stro1989,skru1990,calv2002,espa2007a,espa2007b}.

In addition to the information encoded in the SEDs, grain sizes and composition have been inferred from mid-infrared emission features formed by the super-heated dust grains located at small radii and in the upper atmospheres of these protoplanetary disks.  High-sensitivity, low-resolution spectral surveys of the 10~$\mu$m silicate complex have provided the opportunity for in-depth analyses of compositional variations of the dust grains with the most detailed models including laboratory measured opacities for a variety of sizes and species \citep[see e.g.][]{bouw2003,meeus2003,hond2006,sche2006,sarg2009}.  The parameters returned by these models include dust temperatures and mass fractions for different grain types, sizes, and structures (i.e., crystalline and amorphous, small and large, forsterite, enstatite, olivine).  Less detailed compositional analyses have been conducted by binning the 10~$\mu$m silicate complex into various wavelength regions corresponding to the peak emission of specific dust grain species.  The integrated fluxes over these wavelength bins are used to compute ratios commonly referred to as crystalline indices that give an approximation of the relative abundances \citep[see e.g.,][]{bouw2001,przy2003,vanb2003,kess2005,kess2006,wats2009}.

Several groups have published single-epoch \emph{Spitzer} IRS surveys of TTS in which they note source-to-source morphological differences of the silicate feature \citep{forr2004,kess2005,furl2006,sarg2009,wats2009}.  \citet{forr2004} were the first to note a relationship between the crystallinity versus amorphous nature of the silicate emission feature with the evolutionary status of the dust grain population and the disk.  The presence of crystalline grains implies that the once amorphous interstellar grains have been annealed and cooled, consistent with heating in the inner regions of an evolving protoplanetary disk.  \citet{kess2005} and \citet{sarg2009} concluded the same, suggesting that a disk evolutionary sequence can explain the variety of shapes observed for the silicate features in their Short-Low (5.2-14.5~$\mu$m) and Long-Low (14.0-38.0~$\mu$m) IRS spectra of TTS.

The conclusions about the evolutionary status of the disks are based solely on single-epoch observations and do not account for the dynamic nature of the TTS sources, which are characterized by multi-wavelength variability on timescales as short as days.  Variations of the morphology of the silicate emission feature on the timescales of days, months, or even years would likely complicate such compositional studies of dust grain populations such as those leading to the development of a disk evolutionary sequence.

In order to test the validity of the conclusions drawn from the single-epoch studies, we used \emph{Spitzer} IRS to conduct a spectroscopic variability study of the 10~$\mu$m silicate emission feature in 11 actively accreting classical TTS found in the Taurus--Auriga star forming region.  In 2 of the 11 sources, DG~Tau and XZ~Tau, significant variability of the 10~$\mu$m feature is observed.  The unexpected short timescale variations and potential scenarios for this behavior are presented in this Letter.  A long paper presenting more detailed modeling of these features and the spectra of the sources with no detectable variability is forthcoming (Leisenring et al., in preparation).

\begin{deluxetable*}{crcccrrr}
\tabletypesize{\scriptsize}
\tablecaption{Observations, Continuum and 10~$\mu$m Feature Measurements \label{tbl:1}}
\tablewidth{0pt}
\tablehead{
				\colhead{Star}
			&   \colhead{AOR}
			&  	\multicolumn{2}{c}{Date Observed}
			&  	\colhead{SL1 \& SL2}
			&   \colhead{Mean Slit}
			&  	\colhead{$[6.0]-[13.5]$}
			&  	\colhead{Silicate Flux}  \\
				\colhead{Name}
			&   \colhead{ID} 
			&   \colhead{Gregorian}
			&	\colhead{Julian}
			&	\colhead{Cycles $\times$ Ramps}
			&   \colhead{Offset ($^{\prime\prime}$)}
			&	\colhead{Color\tablenotemark{a}}
			&	\colhead{(10$^{-14}$ W m$^{-2}$)\tablenotemark{b}}}
\startdata
DG Tau & 3530496 & 2004-03-02 & 2453067 & 2 $\times$ 6 sec & -1.15 & 2.63 $\pm$ 0.007 & 1.93 $\pm$ 0.370 \\
\nodata & 15110144 & 2005-10-09 & 2453653 & 8 $\times$ 6 sec & -1.97 & 2.80 $\pm$ 0.009 & 4.05 $\pm$ 0.908 \\
\nodata & 15115008 & 2005-10-15 & 2453659 & \nodata & -2.15 & 2.87 $\pm$ 0.009 & 3.99 $\pm$ 0.866 \\
\nodata & 15115264 & 2006-03-17 & 2453812 & \nodata & 0.00 & 2.90 $\pm$ 0.011 & 6.24 $\pm$ 0.076 \\
\nodata & 15115520 & 2006-03-22 & 2453817 & \nodata & 0.00 & 2.90 $\pm$ 0.011 & 6.00 $\pm$ 0.077 \\
\nodata & 19488000 & 2006-10-17 & 2454026 & \nodata & -0.23 & 2.66 $\pm$ 0.010 & 5.30 $\pm$ 0.057 \\
\nodata & 19487744 & 2006-10-22 & 2454031 & \nodata & -0.40 & 2.64 $\pm$ 0.008 & 3.98 $\pm$ 0.166 \\
\nodata & 19487488 & 2007-03-21 & 2454181 & \nodata & 0.23 & 2.75 $\pm$ 0.010 & 5.78 $\pm$ 0.102 \\
\nodata & 19487232 & 2007-03-28 & 2454188 & \nodata & 0.96 & 2.85 $\pm$ 0.011 & 6.52 $\pm$ 0.516 \\
XZ Tau & 3531776 & 2004-03-04 & 2453069 & 2 $\times$ 6 sec & -0.30 & 2.35 $\pm$ 0.008 & 4.43 $\pm$ 0.163 \\
\nodata & 15111680 & 2005-09-12 & 2453626 & 8 $\times$ 6 sec & 0.46 & 2.37 $\pm$ 0.006 & 1.29 $\pm$ 0.087 \\
\nodata & 15118336 & 2006-03-09 & 2453804 & \nodata & 0.36 & 2.42 $\pm$ 0.007 & 4.16 $\pm$ 0.090 \\
\nodata & 15118592 & 2006-03-17 & 2453812 & \nodata & 0.66 & 2.44 $\pm$ 0.009 & 4.04 $\pm$ 0.212 \\
\nodata & 16875776 & 2006-03-21 & 2453816 & \nodata & 0.75 & 2.42 $\pm$ 0.006 & 3.67 $\pm$ 0.238 \\
\nodata & 19494144 & 2006-10-17 & 2454026 & \nodata & -0.32 & 2.50 $\pm$ 0.007 & 6.12 $\pm$ 0.094 \\
\nodata & 19493888 & 2006-10-22 & 2454031 & \nodata & -0.57 & 2.50 $\pm$ 0.007 & 5.86 $\pm$ 0.236 \\
\nodata & 19493632 & 2007-03-22 & 2454182 & \nodata & 0.26 & 2.34 $\pm$ 0.010 & 3.92 $\pm$ 0.066 \\
\nodata & 19493376 & 2007-03-28 & 2454188 & \nodata & 0.10 & 2.34 $\pm$ 0.008 & 3.97 $\pm$ 0.046 \\ 
\enddata

\tablenotetext{a}{Errors for the color temperature are based off of relative flux uncertainties.}
\tablenotetext{b}{Errors for the flux of the 10~$\mu$m silicate feature are based off of pointing-induced flux uncertainties.}
\end{deluxetable*}

\section{Observations and Data Processing}

Low resolution ($R$~$\sim$~100) spectra of our selected sample of actively accreting TTS were obtained using IRS in Short-Low mode (3\ptsec6 slit and 5.2-14.5~$\mu$m wavelength coverage) in \emph{Spitzer} Cycles~2 and 3.  In order to probe silicate variability on week-, month-, and year-long timescales, each source was observed twice in each of the two approximately month-long visibility windows available per \emph{Spitzer} Cycle.  As a result, we collected a total of 88 spectra spanning a baseline of 1.5 years\footnote{For a couple of sources, only one observation was made in the first visibility window of Cycle~2 and three were made in the other for a given cycle.}.  Archival IRS observations have been included to extend the temporal coverage to a longer timescale.  Table~\ref{tbl:1} summarizes the multiple epochs of observations of DG~Tau and XZ~Tau and includes measurements for the continuum color and the integrated flux of the silicate complex.

\begin{figure}
\includegraphics[angle=90,width=1.0\columnwidth]{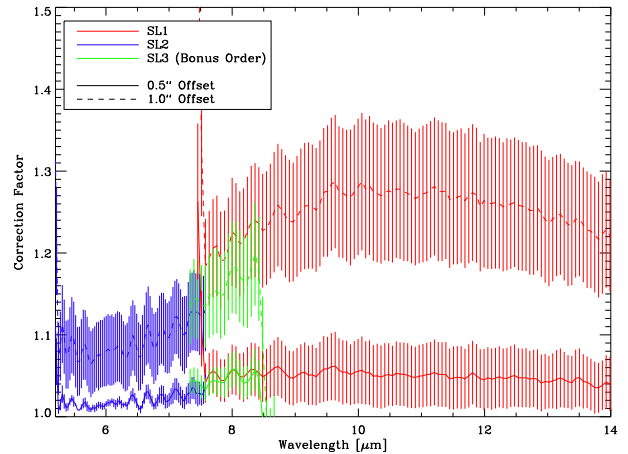}
\caption{Examples of the wavelength-dependent correction factors employed to account for flux loss due to the target being offset from the center of the slit by 0\ptsec5 (solid line) and 1\ptsec0 (dashed line).   The blue portion and red portion of the spectra corresponds to Short-Low order two (5.2-7.7~$\mu$m) and one (7.4-14.5~$\mu$m), respectively.  The green portions of the spectra correspond to the bonus order segment (7.3-8.7~$\mu$m).  Error bars are based on a pointing accuracy of 0\ptsec2 (SSC Helpdesk, private communication) and represent the standard deviation in the absolute flux corrections to the spectra and not errors in flux from one wavelength resolution element to the next.}
\label{fig:pointing}
\end{figure}

The spectral images were reduced using the SMART software package created by the IRS instrument team \citep{higd2004}, the Basic Calibrated Data (BCD) produced by the IRS data pipeline, and routines standardized by the Guaranteed Time Observer IRS Disks team \citep[e.g.,][]{sarg2006}.  Due to the broad wavelength coverage of IRS, the shapes and fluxes of the continuum and solid-state features can be affected by variations of the pointing of the telescope for different nod positions and from epoch to epoch.  Using an average of well-pointed staring observations (AORs: 16344064, 19438080, 20957184, 22165760, 24240896, 24574464, 27392768, and 28709632) and spectral maps (AORs: 16295168 and 19324160) of the standard star HR~7341, the slit offsets were reconstructed and the science spectra were corrected for any epoch-to-epoch mispointings\footnote{The correction factors applied to each spectrum, along with the details of the reconstructed pointing measurements will be included in a Leisenring et al.\ (in preparation).}.  After applying the correction factors corresponding to the appropriate mean slit offsets (see Figure~\ref{fig:pointing} and Table~1, Column 6), we confidently rule out slit offsets as a source for the variations observed in the shapes and fluxes for the silicate features reported in Section 3.  In addition, given the size of the slit (3\ptsec6 for IRS), slit offsets, and the 1\ptsec2 to 3\ptsec6 FWHM of the point spread function for \emph{Spitzer} over the relevant wavelength range of 5 to 15~$\mu$m, the spectral variability observed in the XZ~Tau binary system (separation of 0\ptsec3; \cite{haas1990}) is unlikely to be caused by light loss from either of the stellar components.

\begin{figure}
\includegraphics[angle=90,width=1.0\columnwidth]{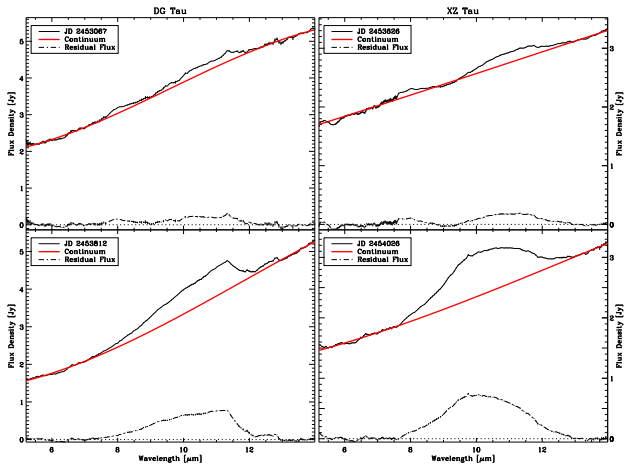}
\caption{Two epochs of spectra for both DG~Tau and XZ~Tau are the solid black lines in each panel.  The smooth gray line represents the best third-order polynomial fit to the short (5.2$-$6.4~$\mu$m) and long (13$-$14~$\mu$m) wavelength portions of the spectra.  For comparison, the residual spectra (black dot-dashed lines) depict the shapes and relative fluxes of the silicate features, clearly illustrating the large variability observed between these epochs of observations.}
\label{fig:example}
\end{figure}

The spectra with continuum fits in Figure~\ref{fig:example} illustrate the technique employed for removing the continuua from the target star spectra and demonstrate that the observed variability of the silicate feature is independent of variations in the long and short wavelength continuum regions.  In order to quantify the shape of the continuua and underlying dust continuum emission, we defined two wavelength bands centered at 6.0 and 13.5~$\mu$m with widths of 0.8 and 1.0 $\mu$m, respectively, and the color index ($[6.0]-[13.5]$; see Table~1).

\section{Results}
\label{results}
Of the 11 sources included in our program, only DG~Tau and XZ~Tau exhibit variability of the silicate complex over month- and year-long timescales.\footnote{We note that the preliminary data analysis found evidence for short day-to-day and week-long timescale variations \citep{leis2007}.  However, once the correction factors for mispointings were applied, these short timescale variations were no longer confidently detected.}  In both systems, the silicate emission feature is observed to weaken and strengthen over the three years worth of observations, with the short wavelength portion near 9~$\mu$m in XZ~Tau falling nearly to the level of the continuum.  The silicate variability in the spectrum of DG~Tau has previously been observed \citep{wood2000,wood2004,sitk2008} and is confirmed by these observations, while the variations in the spectra of XZ~Tau are observed for the first time.

\begin{figure*}
\includegraphics[angle=90,width=2.0\columnwidth]{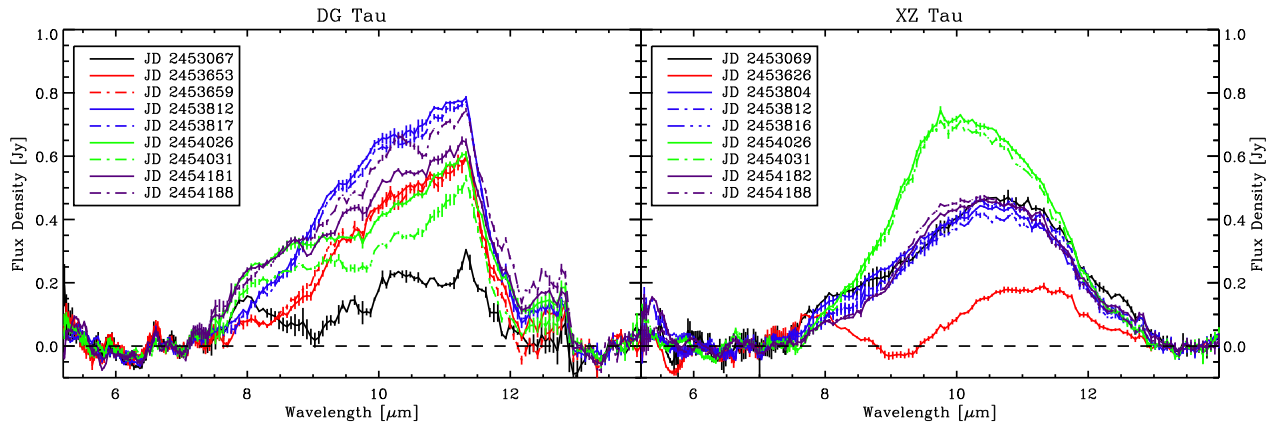}
\caption{All epochs of spectra acquired for both DG~Tau and XZ~Tau with \emph{Spitzer} IRS.  In each case, the silicate feature represents a different compositional signature.  The silicate feature for DG~Tau appears strongly peaked near 11.3~$\mu$m the location of crystalline forsterite emission, while the very smooth silicate feature of XZ~Tau peaks at shorter wavelengths and lacks this notable crystalline emission.  Therefore, by inspection and the previously used crystalline indices, the silicate dust grain populations are characterized by an abundance of crystalline dust grains in the disk of DG~Tau and amorphous grains for XZ~Tau.  Note that the absolute flux error bars show that the small variations are not likely to be real, which rule out week-long silicate variability in these observations.}
\label{fig:silew}
\end{figure*}

Using the JD~2453069 observation of XZ~Tau, \citet{sarg2009} determined that the silicate feature is dominated by amorphous grains by fitting a two-temperature blackbody with solid angle dependence and laboratory measured opacities for a variety of dust grain sizes and compositions.  By inspection of the XZ~Tau spectra (Figure~\ref{fig:silew}), for all epochs, this source lacks evidence of emission from crystalline forsterite at 11.3~$\mu$m, a common signature of sources determined to possess relatively high abundances of crystalline grains \citep{forr2004,kess2006,sarg2009}.  On the other hand, the silicate emission feature for DG~Tau appears dominated by crystalline forsterite in all epochs of observations presented here\footnote{\citet{wats2009} were unable to confidently extract a meaningful crystalline index from a single-epoch observation for this source.}.  Detailed modeling of the silicate feature is required to extract the crystalline-to-amorphous ratio and will be presented in Leisenring et al.\ (in preparation).

\begin{figure}
\includegraphics[angle=90,width=1.0\columnwidth]{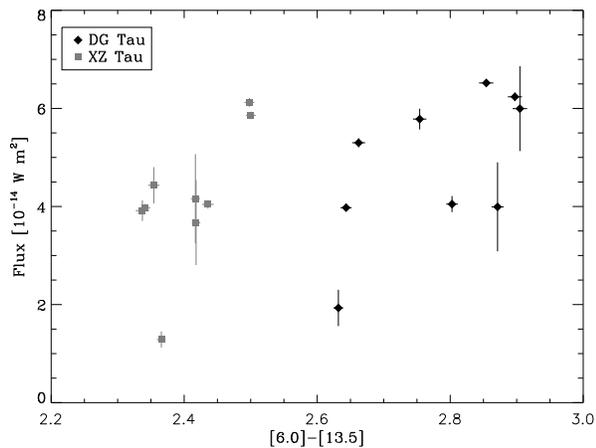}
\caption{Silicate flux plotted vs.\ the $[6.0]-[13.5]$ color for both DG~Tau and XZ~Tau. We obtain respective Pearson correlation coefficients between these two parameters of 0.61 and 0.66 for DG~Tau and XZ~Tau. The probabilities that these correlations could have been produced from a random distribution \citep[see, for example,][]{tayl1997} are 8.1\% and 5.3\%, respectively. Thus, we find only a weak correlation between the strength of the silicate feature and the continuum flux.}
\label{fig:colorflux}
\end{figure}

In Figure~\ref{fig:colorflux}, the silicate flux appears to be marginally correlated with the  $[6.0]-[13.5]$ color index for both DG Tau and XZ Tau. However, such weak correlations do not provide compelling evidence for a common physical mechanism that is directly responsible for the variations in the silicate feature and also for changes in continuum emission. We hope to further inquire into this relation with (near-)simultaneous multi-wavelength observations. 

\subsection{Preliminary Modeling}
\label{prelim}

In addressing the evolutionary sequence previously proposed as an explanation for the amorphous and crystalline silicate abundances, we adopted the two-temperature dust models (including the same dust species and related opacities) of \cite{sarg2009} (see Equation~1) and performed compositional fits to the 10~$\mu$m silicate feature at each phase of modulation for DG~Tau and XZ~Tau.  The models returned statistically significant fits only for the epochs possessing maximum silicate emission (DG Tau: JD 2453653 and 2453659; XZ Tau: JD 2454026 and 2454031).  Therefore, the models were unable to provide a reliable fit to the silicate feature when the emission was diminished or self-absorbed.  The failure of this simple method led us to develop a two-temperature two-slab model of the dust emission from the surface of the disk.  In order to produce the observed variations of the silicate feature, the top, intervening layer was assumed to be cooler than that of the lower, emitting region.  This two-slab model confidently fits all epochs of the silicate emission features for both sources, especially when both slabs possessed dust grains with similar compositions.  The results of this model\footnote{The models and results will be detailed in Leisenring et al.\ (in preparation).} suggests that the observed variability can be explained by an intervening population of similarly composed cooler dust grains.  Therefore, the variations observed in the silicate feature do not necessarily indicate a change in the composition of the emitting grains.  The introduction of an absorbing column of dust grains explains the inability of previous silicate decomposition models to fit silicate emission features in DG~Tau and other similar sources \citep{sarg2009,wats2009}.

\section{Intervening Cooler Dust Grain Scenarios}

The timescales over which the variations occur strongly suggest that the phenomena associated with intervening  cooler dust must occur on dynamical timescales consistent with the motion of dust in the disk at radii less than or equal to 1~AU.  These short timescales rule out a clumpy dust envelope as the source for the intervening column of cooler dust.  This inference agrees with the findings of \cite{wood2004}, in which a one-dimensional radiative transfer model showed that an envelope capable of producing the variability they observed in DG~Tau would also completely obscure the star.  Therefore, we believe that it is unlikely that the intervening cooler grains are associated with remnants of an envelope.

In an attempt to explain day- to week-long silicate and continuum variability, \citet{muze2009} discuss a scenario of disk shadowing in which a warped, puffed up region of the inner disk rim casts a shadow across a portion of the disk surface.  There is potential for this phenomena to explain the longer timescale variability presented here \citep{sitk2008}.  In addition, the recent discovery of a third companion located within 0\ptsec09 (13~AU) of the southern component of XZ~Tau \citep{carr2009} increases the possibility for variations in disk illumination as previously described by \citet{wats2007} in reference to the variability they observe in the HH~30 system.

In addition to disk shadowing and illumination, turbulent mixing \citep{turn2009, bals2008} and the presence of dusty disk winds may also provide reasonable explanations of the observed variability.  Turbulent mixing in the disk capable of dredging cooler dust up to the surface of the disk may provide a plausible mechanism for varying the strength and nature (emission versus self-absorption) of the silicate feature.  Alternatively, the presence of dusty disk winds as described by \cite{tamb2008} may be capable of lofting material above the disk surface providing a cooler veil of similarly composed dust grains through which to view the warmer, emitting dust; hence, producing self-absorption of the emitting silicate region.  In both scenarios, the vertical mixing and disk winds must be variable to produce the inhomogeneities required to explain the observed variability of the silicate feature.  Given the variable nature of the accretion processes in these sources, fluctuating disk winds and vertical mixing is to be expected \citep{bary2008,nguy2009}.

\section{Conclusions}

The 10~$\mu$m silicate emission complex observed in the mid-infrared spectra of classical TTS and formed by super-heated dust grains in the upper atmospheres of the innermost regions of their disks can vary on month-long timescales.   Better temporal resolution is required to determine if the variability is periodic.  Our spectral decomposition models were incapable of fitting the silicate feature at every phase of modulation, suggesting that variability may be an explanation for some of the difficulties previous groups have encountered with modeling single-epoch spectra of individual stars.  The simple two-slab two-temperature model suggests an optical depth effect may explain the observed variations.  Since the intervening cooler material may be of similar composition, the variability may not directly impact the disk evolutionary sequence previously constructed for TTS, based on amorphous versus crystalline abundances and dust grain sizes.  However, we note that these models require a number of parameters and the question of degenerate solutions needs to be addressed.  Additional multi-epoch observations of these sources and others are required to determine the fraction of disks with variable silicate emission and provide more variations to model, helping to constrain acceptable parameters.  With the loss of IRS on \emph{Spitzer}, this project will turn to ground-based observations with less sensitivity, but still the ability to measure the previously observed silicate variability.

\acknowledgements

The authors thank Gregory Herczeg, Lynne Hillenbrand, Ciska Kemper, James Muzerolle, Dawn Peterson, and Mike Sitko for useful discussions.  JSB acknowledges support through a NSF Astronomy and Astrophysics Postdoctoral Fellowship grant AST 05-07310.  JSB and JML acknowledge support through \emph{Spitzer} Cycle 2 and 3 grants RSA 1277404 and 1288859.



\end{document}